\documentclass[twocolumn]{emulateapj}
\usepackage{multirow}
\usepackage{graphics}
\usepackage{amsmath, amsthm, amssymb}
\usepackage{epsfig}
\usepackage[hidelinks]{hyperref}
\usepackage{natbib}
\usepackage{float}
\usepackage{lineno}

\usepackage{graphicx}
\usepackage[dvipsnames]{xcolor}

\begin{document}

\title{
Comparing dynamical effects of the central bar and the spiral arms in the solar neighborhood
}

\author{Willian Nacafucasaco$^{1}$}\email[e-mail: ]{willianyn@usp.br}
\author{Tatiana Michtchenko$^{1}$}\email[e-mail: ]{tatiana.michtchenko@iag.usp.br}
\author{Douglas Barros$^{2}$}\email[e-mail: ]{douglas.barros@alumni.usp.br}
\author{Jacques Lépine$^{1}$}\email[e-mail: ]{jacques.lepine@iag.usp.br}

\affiliation{$^1$Universidade de S\~ao Paulo, IAG, Rua do Mat\~ao, 1226, Cidade Universit\'aria, 05508-090 S\~ao Paulo, Brazil\\
$^2$Escola T\'ecnica Estadual Professor Agamenon Magalh\~aes, Avenida Jo\~ao de Barros, 1769, Espinheiro, 52021-180, Recife, Brazil}

\date{\today}

\begin{abstract}
    The dynamical effects on the stellar motion produced by the Galactic central bar and the spiral arms perturbations are investigated separately and compared. The stars from the \textit{Gaia} DR3 catalog are selected in the region of observable completeness, which we estimate as $\sim$1\,kpc from the Sun. We apply the 2D model of the Galactic potential consisting of three axisymmetric components, the disk, the bulge, and the dark matter halo, and two non-axisymmetric components, the central bar and the spiral arms.  The stellar dynamics is studied using analytical and numerical techniques, such as Hamiltonian topology analysis, the construction of dynamical maps on the representative planes, dynamic spectra, and Poincaré sections.  We identify the main dynamical features in the solar neighborhood (SNd), the corotation (CR) and Lindblad resonances (LRs). By assuming that the main moving groups (MGs) in the SNd originate from the resonances, we compare their locations, structures, and intensities with the theoretical predictions and provide a description of the process involved in the formation of the MGs. In addition, we explore parametric planes by adjusting the values of the pattern rotation speed $\Omega_{p}$ with the positions of the MGs, for both the spiral arms and bar models, and conclude that the spiral arms model shows better results when compared to those of the bar, under the hypothesis of the dynamical origin of MGs.

   \end{abstract}


\maketitle


\section{Introduction} 
\label{sec:intro}

{The central bar and the spiral arms are key non-axisymmetric structures that modify both the morphology and the dynamical properties of the unperturbed Galactic disk. The perturbations produced by these structures add significant complexity to the dynamics of stars. The prevailing idea is that the distinct stellar structures and groupings observed in the Solar Neighborhood (SNd) may result from resonances and chaotic motions driven by the Galactic bar and spiral arms.  However, there is still no agreement on the respective contribution of each of these two components. Clarifying this issue would help us to better interpret the various observable phenomena in the SNd, particularly the \textit{moving groups} (MGs) in the velocity space around the Sun}.

Today, the consensus is that MGs have the same dynamical origin. The most plausible scenario suggests that these groups arise from stars trapped in resonance zones within the Galactic disk. Some researchers link the formation of specific groups, such as the Hercules Moving Groups (HMG), to resonances driven by the Galactic central bar (\citealp{2000AJ....119..800D}, \citealp{2017ApJ...840L...2P}, \citealp{2019A&A...626A..41M}, \citealp{2020MNRAS.499.2416A}, \citealp{2023ApJ...956..146L}, \citealp{2025MNRAS.538.1963L}). Other studies explain the dynamical origin of these MGs using the spiral arms of the Galaxy (\citealp{2018ApJ...863L..37M}, \citealp{2020ApJ...888...75B}). Furthermore, some researchers explore the more complex case by combining the effects of both structures (\citealp{2020MNRAS.496.1845M}).  In this context, the aim of the present work is to systematically compare the dynamical effects on the stellar motion produced by the Galactic spiral arms and the central bar in the SNd.

{ For this comparative analysis, we use the main MGs and \textit{diagonal ridges} (DRs) as test structures, whose properties in the full phase space are supplied by observational data from \textit{Gaia} DR3 (\citealp{2023A&A...674A...1G}). The velocity-space at the Sun's location perturbed by the bar and the spiral arms, separately, has already been presented and compared in \cite{2018A&A...615A..10M}. Here, we perform a detailed analysis of the topology of the complete 2D phase space of the SNd, which arises from two non-axisymmetric structures.}

{ By assuming the dynamical origin of the MGs and DRs, we apply well-established analytical and numerical techniques widely used in celestial mechanics studies, such as the analysis of the Hamiltonian topology, the construction of the dynamical maps on the representative planes, dynamical spectra and Poincaré sections through the numerical integrations of the stellar motion.}
{The Galactic two-dimensional potential is obtained by fitting the observed rotation curve and by analytically modeling the non-axisymmetric components, following the approach of \cite{2018A&A...615A..10M}. The novelty is the improved representation of the central bar, which is now modeled by three layers of homogeneous concentric ellipsoids of different densities, in this way, fitting a non-uniform matter distribution between the short, middle and long extended structures in the central Galactic region.}

{Our study shows that, for both the spiral arm and central bar configurations, Lindblad resonances (LRs) play a role in shaping the MGs and the DRs, but the effect is dominated by the spiral arms. A detailed analysis of the resonant orbits enables us to elucidate the mechanism of formation of the MGs through the capture and long-term retention of stars within resonance regions.
Considering the strong sensitivity of the LRs configuration in phase space to the pattern speeds, we further extend our analysis across the parametric planes, varying both the bar and spiral arm speeds on large scales, from 20\,km\,s$^{-1}$\,kpc$^{-1}$ to 60\,km\,s$^{-1}$\,kpc$^{-1}$.}

This paper is organized as follows. In Sect.\,2, we describe the sample used in our analysis and introduce our choices of the two representative planes. In Sect.\,3, we introduce our analytical model for both spiral arms and a central bar, in addition to the axisymmetric component. In Sect.\,4, we discuss the choice of the physical parameters used in the model. In Sect.\,5, we show the Hamiltonian topology of the system and introduce the corotation and Lindblad resonances. In Sect.\,6, we present a dynamical view of the MGs on the representative planes and offer a brief analysis of the dynamical environment in the SNd. In Sect.\,7, we discuss the mechanisms that could be responsible for the formation of MG. In Sect.\,8, we study the parametric planes by means of dynamical mapping as a function of the pattern speed. Finally, in Sect.\,9, we summarize our results and present our conclusions.


\section{Data sample and its dynamical features}\label{sec:sample}

Our sample contains stars from \textit{Gaia} DR3 catalog (\citealp{2023A&A...674A...1G}), with astrometry quality measures limited to $ruwe < 1.4$, ensuring accurate positive values of parallax, right ascension, declination and proper motions with good quality. Moreover, we choose all objects for which line-of-sight velocity measurements have an absolute uncertainty of less than 5\%. This value was selected because it seems to be a good compromise between data quality and sample size. Finally, we restricted our sample to include only objects that have more than 8 transits and excluded binary stars.

\begin{figure}
    \includegraphics[width=1.\columnwidth]{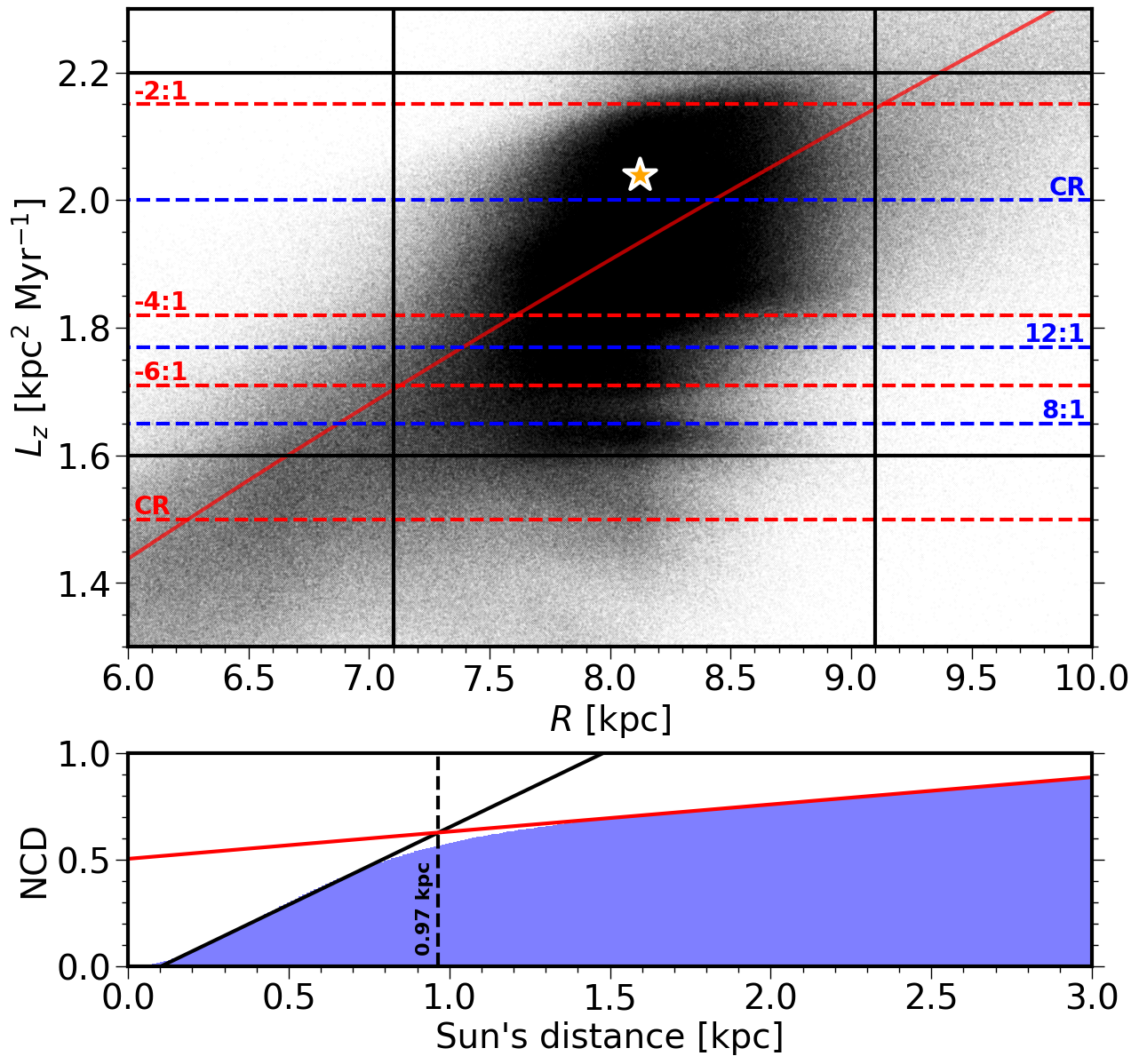}
    \caption{The sample used in this work, on the $R$--$L_z$ plane (top panel), and the Normalized Cumulative Distribution (NCD) of the stars from the sample as a function of the distance to the Sun (bottom panel). \textit{Top}: The assumed limits of the sample's observable completeness are indicated by two vertical ($R$--range between 7.1\,kpc and 9.1\,kpc) and two horizontal ($L_z$--range  between 1.6\,kpc$^{2}$\,Myr$^{-1}$ and 2.2\,kpc$^{2}$\,Myr$^{-1}$) black continuous lines. The blue and red dashed lines indicate the nominal locations of the main LRs, produced by spiral arms and the central bar, respectively (see in Sect.\,\ref{sec:mapping-resonances}). The continuous red curve shows the location of the rotation curve on the $R$--$L_z$ plane, while the star symbol is the location of the Sun on this plane. \textit{Bottom}: The cumulative distribution is normalized by the maximum value, defined at $\sim 6$\,kpc from the Sun. The solid black and red lines are two linear fits of the NCD; their intersection at $\sim 0.97$\,kpc is chosen as a starting point for the  decrease in the accretion rate of the NCD.
}
    \label{fig:RLz-distribution}
\end{figure}

Using right ascension, declination and parallax, we first compute the rectangular spatial coordinates of stars in the heliocentric frame. In the next, adopting the Sun's position at $R_\odot=8.122$\,kpc, $\varphi_\odot = 90$\,deg and $Z_\odot = 15$\,pc, and the velocity vector $\left<-12.9,245.6,7.78\right>$\,km\,s$^{-1}$ (\citealp{2018RNAAS...2..210D}), we calculate the cylindrical coordinates  $\left<R,\varphi,Z\right>$ and velocity components $\left<p_R,V_\varphi,V_z\right>$ of the objects in the Galactocentric reference frame\footnote{
In order to maintain consistency with notation used in this work, we prefer to use the cylindrical coordinates in the Galactocentric frame, instead the rectangular Galactic velocities in the heliocentric frame,  $\left<U,V,W\right>$. It is worth noting that the observational features are invariant by translation and rotation in the position and velocity spaces.
}. All transformations were performed using the \textit{astropy} library in \textit{Python}.

In this paper, we analyze the dynamics of stars in the equatorial Galactic plane by applying a 2D Galactic gravitational potential model in the SNd. In order to compare the theoretical predictions with the observed distribution of stars, we select our sample from the thin disk objects. 
According to \cite{2016ARA&A..54..529B}, the vertical size scale and the star dispersion velocities of the thin disk are within the ranges of $220$--$450$\,pc and $20$--$27$\,km\,s$^{-1}$, respectively. Therefore, we can restrict our sample to stars with $|Z| < 200$\,pc and $|V_z| < 20$\,km\,s$^{-1}$. Moreover, since we focus our attention on the solar vicinity, we constrain the values of the radial and tangential velocities to the ranges of $|p_R| < 80$\,km\,s$^{-1}$ and $V_\varphi = 0$--$350$\,km\,s$^{-1}$, respectively. In this way, we obtain the sample of around 5.7 million objects.

Figure \ref{fig:RLz-distribution} shows on the top panel the sample obtained on the $R$--$L_z$ plane, where $L_z$ is the vertical component of the stellar angular momentum, defined as $L_z=R\times V_{\varphi}$.
{For the sake of reliable application in the comparative analysis, we define a confidence zone, within which we consider our sample to be observably complete. The limits of this zone are indicated by two vertical and two horizontal  black solid lines on the top panel in Fig.\,\ref{fig:RLz-distribution}.}

{To estimate the region of observable completeness, we applied the Normalized Cumulative Distribution (NCD) method.} 
The NCD-curve of stars from our sample is shown on the bottom panel in Fig.\,\ref{fig:RLz-distribution}, as a function of the stellar distance from the Sun. The black and red solid lines are two linear fittings of the evolution of the NCD along the radial distance from the Sun, illustrating a visible decrease in the accretion rate that occurs around 0.97\,kpc (vertical dashed line).
Using this value, we establish the approximate completeness boundaries of our sample, within which the analysis of the stellar distribution is not affected by observational biases. It is worth mentioning that a similar value for the Sun's completeness neighborhood was obtained in \cite{2025arXiv250703742K} by applying distinct criteria.

Finally, from a dynamical point of view, the radial velocity $p_R$ can be set to $p_R = 0$, without loss of generality, as this condition will be encountered by all orbits. In addition, the standard deviation of the variable $\varphi$ from $90$\,deg is only 9.5\,deg in our sample, which justifies fixing $\varphi=90$\,deg.

\begin{figure*}
    \includegraphics[width=1.95\columnwidth]{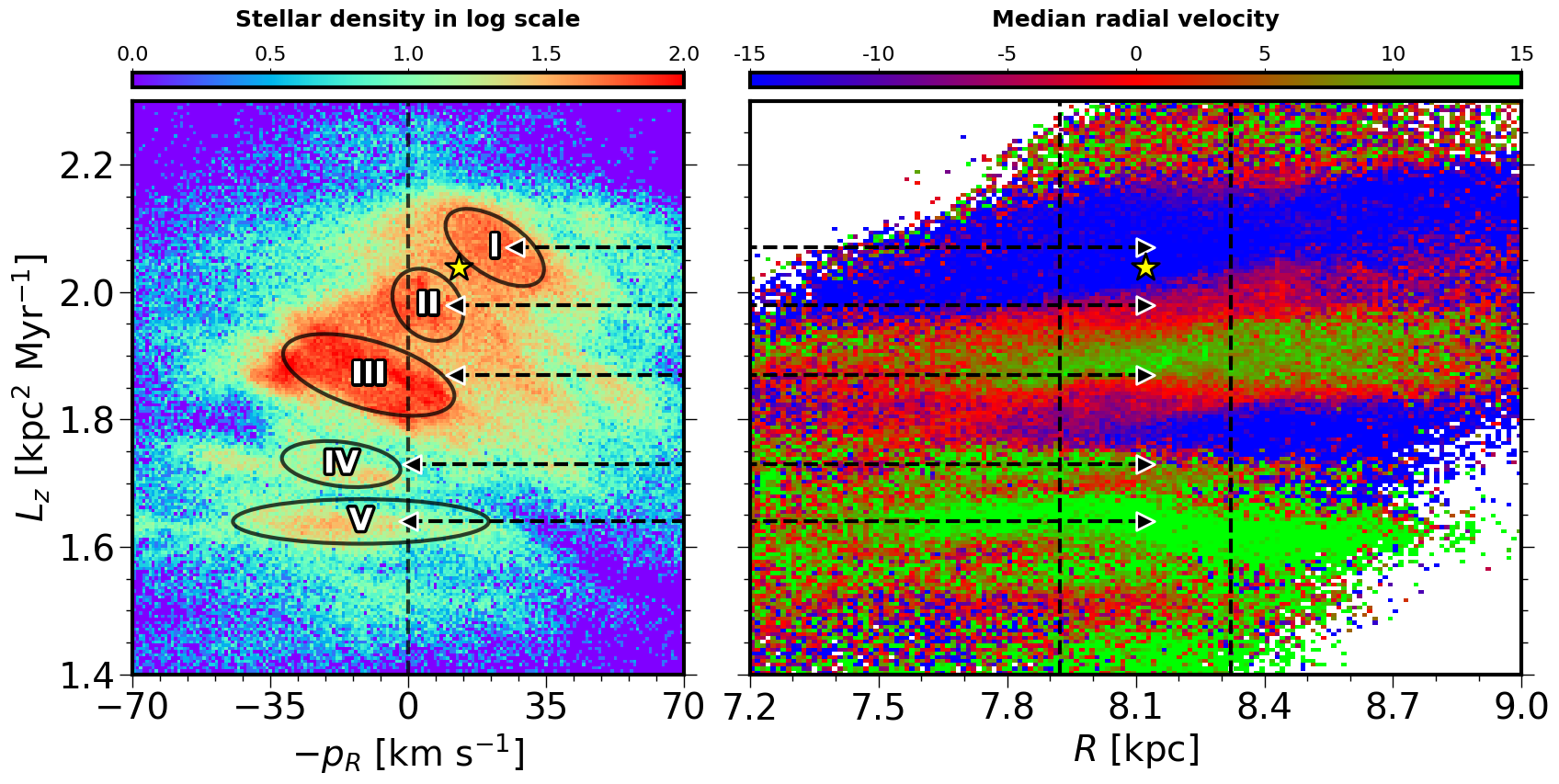}
    \caption{Distribution of the stars on the $p_R$--$L_z$ plane (left panel) and the $R$--$L_z$ plane (right panel). \textit{Left}: The main MGs calculated within $|R-R_\odot| < 200$\,pc, are schematically indicated by black ellipses (I - Sirius, II - Coma Berenice, III - Hyades \& Pleiades, IV - HMG I, V - HMG II). The color bar on the top of the panel shows, in logarithmic scale, the density of the stars calculated in bins 0.8\,km\,s$^{-1}$ $\times$0.005\,kpc$^2$Myr$^{-1}$. \textit{Right}: The DRs calculated in bins {0.012}\,kpc$\times${0.007}\,kpc$^2$\,Myr$^{-1}$ show, in color scale, the distribution of the median of radial velocity $p_R$. 
    The vertical dashed lines delimit the vicinity of the Sun inside $200$\,pc. Finally, the horizontal arrows set up a correspondence between the diagonal ridges and the MGs, while the star symbols in both panels shows the position of the Sun in these planes for reference.}
    \label{fig:moving-groups}
\end{figure*}

\subsection{Moving groups and diagonal ridges}\label{sec:representative-planes}

Figure \ref{fig:moving-groups} shows the classical MGs (left panel; {see \citealp{2014A&A...563A..60A}}) and the DRs (right panel), calculated with objects from our sample. The density distribution of the sample stars from the region within the distance of 200\,pc from the Sun is shown in the momentum space $p_R$--$L_z$ ($p_R$ and $L_z$ being the radial velocity and the vertical component of the angular momentum, respectively). The phase plane $R$--$L_z$ (right panel) shows the distribution of the median value of the radial velocity $p_R$, calculated in bins of {0.012}\,kpc$\times${0.007}\,kpc$^2$\,Myr$^{-1}$. Using the same variable $p_R$ to construct both representative planes, we establish a link between them for the sake of comparison. Indeed, the horizontal dashed arrows connect the substructures of the MGs near the Sun (left panel) to the DRs (right panel), which extend well beyond the Sun's proximity. Thus, we can assume that the two planes in Fig.\,\ref{fig:moving-groups} represent a same phenomenon, produced due to gravitational effects of the bar and/or spiral arms; we analyze the dynamical maps on these planes in the following sections.


\section{Galactic potential  model}
\label{sec:model}

Here, we briefly describe the 2D model for the Galactic potential and refer the reader to \cite{2017A&A...597A..39M, 2018A&A...615A..10M} for more details. The model considers the contribution of several axisymmetric components, such as stellar and gaseous disks, a central bulge and a dark matter halo, whose physical and structural parameters are adjusted by the observed rotation curve. In addition, there are two non-axisymmetric components, the four-arm spiral pattern and the central bar, which are represented through analytical modeling. 

\subsection{Axisymmetric component of the Galactic potential}
\label{subsec:axisymmetric}

The rotation curve fitting procedure is described in detail in \cite{2021FrASS...8...48B}. 
Briefly, the rotation curve is expressed as 
\begin{align}\label{eq:Vrot}
    V_{\textrm{rot}}(R) &= a_1\exp\left[-\frac{R}{a_2} - \frac{a_3}{R} \right] \nonumber \\
                        &\hspace{2.cm}+ a_4\exp\left[-\frac{R}{a_5} - \left(\frac{a_6}{R}\right)^2 \right],
\end{align}
where the coefficients $a_i$ ($i=1,...,6$) are obtained through fitting of the measurements of HI (\citealp{1978A&A....63....7B} and \citealp{1989ApJ...342..272F}) and CO (\citealp{1985ApJ...295..422C}) lines and \textit{masers} associated with high-mass star-forming regions (\citealp{2017AstBu..72..122R} and \citealp{2019ApJ...885..131R}). Table\,\ref{tab:rotation-curve-params} shows the $a_i$--values used in this work. 

\begin{table}
\caption{Fitting parameters of the rotation curve.}
\label{tab:rotation-curve-params}      
\centering   
\begin{tabular}{l c c c}
\hline\hline 
Symbol  &   Value  &   Unit  \\
[0.5ex]
\hline
             $a_1$ & 293 & km\,s$^{-1}$  \\
             $a_2$ & 5.1 & kpc  \\
             $a_3$ & 0.032 & kpc  \\
             $a_4$ & 216 & km\,s$^{-1}$  \\
             $a_5$ & 100000 & kpc  \\
             $a_6$ & 3.8 & kpc  \\
[0.5ex]  
\hline\hline
\\
\end{tabular}
\end{table}

Using the expression of the observable rotation curve, we can calculate the axisymmetric part of the Galactic potential $\Phi_0$ through the integration of the equation
\begin{equation}\label{eq:PotAxs}
    \frac{\partial \Phi_0}{\partial R} = \frac{V^2_{\textrm{rot}}(R)}{R},
\end{equation}
with the boundary condition $\lim_{R\to\infty}\Phi_0(R) = 0$.

\subsection{Spiral arms potential}

The spiral arms model that describes the density profile observed in the Galaxy disk was proposed in \cite{2013A&A...550A..91J}. The 2D potential in the equatorial plane of the disk can be written as
\begin{equation}\label{eq:PhiArms}
    \Phi_{\textrm{sp}}(R,\varphi) = -\zeta_0Re^{-g(R,\varphi)},
\end{equation}
where $g(R,\varphi)$ is a function defined by
\begin{equation}\label{eq:gfunction}
    g(R,\varphi) = \frac{R^2}{\sigma^2}\Big[1 - \cos\big(m\varphi - f_m(R)\big)\Big] + \epsilon_sR.
\end{equation}
Here, $f_m(R)$ is a \textit{shape function} that draws the spiral arms in logarithmic format as
\begin{equation}\label{eq:shapefunction}
    f_m(R) = \frac{m}{\tan(i)}\ln\left(\frac{R}{R_i}\right) + \gamma_{\textrm{sp}}.
\end{equation}

The values of the parameters in Eqs.\,\ref{eq:PhiArms}--\ref{eq:shapefunction} are listed in Table\,\ref{tab-init-2} and their choice is discussed in Sect.\,\ref{sec:spiral-parameters}.

\subsection{Central bar potential}

In this work, the central bar with mass $M_{\textrm{bar}}$ is modeled by layers of homogeneous concentric ellipsoids, with a rotation speed of $\Omega_{\textrm{bar}}$ and a phase angle $\gamma_{\textrm{bar}}$ with respect to the reference direction ($X$-axis). The gravitational potential produced by an ellipsoid at a point $P = \left<X,Y\right>$ in the Galactic equatorial plane is given as
\begin{equation}\label{eq:PotElipsoide}
    \Phi(P) = -\frac{3}{2}GM\Big[U_0(\zeta) + X^{*2}U_1(\zeta) + Y^{*2}U_2(\zeta)\Big],
\end{equation}
where $G$ is the Gravitational constant and $M$ is the mass of the layer. The set $\left<X^*,Y^*\right>$ defines the bar's reference frame coordinates calculated as 
\begin{align*}
    X^* &= R\cos(\varphi - \gamma_{\textrm{bar}}), \\
    Y^* &= R\sin(\varphi - \gamma_{\textrm{bar}}).
\end{align*}

In the case of the elongated ellipsoids, the coefficients $U_i$ ($i=0,1,2$) are analytical functions of the variable $\zeta$ (see \citealp{2018A&A...615A..10M} for more details):
\begin{align}
    U_0(\zeta) & = \frac{1}{c\zeta_0}\ln\left(\sqrt{1 + \zeta^2} + \zeta\right),\label{eq:BarraU0} \\
    U_1(\zeta) & = \frac{1}{c^3\zeta_0^3} \left[\frac{\zeta}{\sqrt{1 + \zeta^2}} - \ln\left(\sqrt{1 + \zeta^2} + \zeta\right)\right],\label{eq:BarraU1} \\
    U_2(\zeta) & = \frac{1}{2c^3\zeta_0^3}\left[\ln\left(\sqrt{1 + \zeta^2} + \zeta\right) - \zeta\sqrt{1 + \zeta^2}\right]\label{eq:BarraU2},
\end{align}
where $c$ is the minor semi-axis of the ellipsoid and the variable $\zeta$ is defined as:
\begin{equation}\label{eq:ZetaDefinition}
    \zeta^2 = \frac{a^2 - c^2}{c^2 + \lambda},
\end{equation}
where $a$ is the major semi-axis of the ellipsoid and $\zeta_0$ is computed for $\lambda = 0$. At the point $P$ lying outside the layer, $\lambda$ is the positive square root of the quadratic equation 
\begin{equation}\label{eq:ElipsoideParametric}
    \frac{X^{*2}}{a^2 + \lambda} + \frac{Y^{*2}}{c^2 + \lambda} = 1,
\end{equation}
while, at any point $P$ inside the layer, $\lambda=0$.
It is worth mentioning that the ellipsoid is a 3D figure, but here we take into account only the potential components in the plane $Z = 0$. 

In our investigations, we assumed a three-layer configuration of the central bar with three different densities. This choice is justified by observational evidence that suggests a non-uniform density distribution between the \textit{short}, \textit{middle} and \textit{long} {extended} structures in the central Galactic region (e.g., \citealp{2016ARA&A..54..529B}). The resulting potential of the bar $\Phi_\textrm{bar}$  at a given point is just a sum of the contributions of the inner layer (core ellipsoid), the middle and outer layers at this point. The potential of the middle (outer) layer is just a potential given by Eq.\,(\ref{eq:PotElipsoide}) after subtracting the potential of the inner (middle) component. It should be noted that the axisymmetric contribution of the bar potential has already been taken into account in the observation-based $\Phi_0$. Therefore, we subtract a spherical component from $\Phi_\textrm{bar}$, as described in \cite{2018A&A...615A..10M}; {the resulting mass of the non-axisymmetric part of the bar} is shown in Table\,\ref{tab-init-2}.

\begin{table}
\caption{Physical and structural parameters of the spiral arms and the three-layer central bar of the Galactic potential model.}
\label{tab-init-2}      
\centering   
\renewcommand{\thefootnote}{\arabic{footnote}}
\begin{tabular}{l c c c}
\hline\hline 
Spiral arms &  Symbol  &   Value  &   Units  \\
[0.5ex]
\hline
    Pitch angle\footnote{{It is worth noting our choice of the negative sign of the pitch angle, which means the counterclockwise Galactic rotation from the viewpoint of an observer located towards the direction of the
North Galactic Pole. }} & $i$ & -14.0 & deg \\
    Perturbation amplitude & $\zeta_0$ & 150.0 & km$^2$s$^{-1}$kpc$^{-1}$  \\
    Number of spiral arms & $m$ & 4 &  - \\
    Arms width & $\sigma$ & 8.0 & kpc \\
    Scale size & $\epsilon_s^{-1}$ & 4.0 & kpc  \\
    Reference radius & $R_i$ & 8.122 & kpc \\
    Spirals pattern speed & $\Omega_{\textrm{sp}}$ & 28.0 & km s$^{-1}$ kpc$^{-1}$ \\
    {Initial phase angle} & $\gamma_{\textrm{sp}}$ & 223.34 & deg \\
[0.5ex]  
\hline\hline
Central bar & Symbol & Value & Unit \\
[0.5ex]
\hline
    Bar-Bulge mass & $M_b$ & 2.57$\times 10^{10}$ & $M_\odot$ \\
    {Non-axisymmetric mass} & $M_{\textrm{bar}}$ & $5.33\times10^{9}$ & $M_\odot$ \\ 
    Major semi-axis size & $R_{\textrm{bar}}$ & 4.5 & kpc \\
    Flattening & $f_{\textrm{bar}}$ & 0.7 & - \\
    Bar's pattern speed & $\Omega_{\textrm{bar}}$ & 40.0 & km s$^{-1}$ kpc$^{-1}$ \\
    {Initial phase angle}\footnotemark[\value{footnote}] & $\gamma_{\textrm{bar}}$ & 115.0 & deg \\
[0.5ex]  
\hline\hline
{Bar's layers}  & Inner & Middle & Outer\\
\hline
    Semi-major axis (kpc)     & 2.0 & 2.9 & 4.5 \\
    Mass ($10^{10} M_\odot$) & 0.85 & 0.85 & 0.85 \\
    Density (G$M_\odot$) & 0.0036 & 0.0012 & 0.0003 \\
[0.5ex]
\hline\hline
\end{tabular}
\end{table}


\section{Setting the physical and structural parameters}\label{sec:parameters}

To identify the dynamical effects of the non-axisymmetric structures within the observable stellar distribution, it is necessary to evaluate their physical and structural parameters in the Galactic potential model. The parameters for the spiral arms model were obtained by the observation data fitting procedure, while the parameters for the central bar expressions were selected from the literature. The choice (see Table\,\ref{tab-init-2}) is briefly discussed in this section. 

\subsection{Spiral arms parameters}\label{sec:spiral-parameters}

The spiral arms model has already been introduced and explored in our research series (e.g., \citealp{2017A&A...597A..39M, 2017ApJ...843...48L, 2020ApJ...888...75B}). The values of the structural parameters were derived through the modeling of the stellar spiral density in the Galactic disk and calculated under the self-consistency condition (see \citealp{2021FrASS...8...48B} for details). In particular, the choice of the four-arm spiral model is favored by the spiral tracer observations, as commented on \cite{2017ApJ...843...48L}. 

{The choice of the angular {rotation }speed $\Omega_{\textrm{sp}}$ is very important because the position of the dynamical characteristics induced by the spiral arms is uniquely defined by this parameter. }
{Assuming a rigid body rotation of the spiral arms structure, we use $\Omega_{\textrm{sp}} = 28.0$\,km\,s$^{-1}$\,kpc$^{-1}$ that is consistent with the values derived in \cite{2019MNRAS.486.5726D} and \cite{2022AstL...48..568B}.  Note that our choice of the $\Omega_{\textrm{sp}}$--value is in good agreement with the result in \cite{2025arXiv250703742K}, where the authors model the distribution of stars in the SNd and obtain the best-fit for the pattern rotation speed of 28.2$\pm$0.1\,km\,s$^{-1}$\,kpc$^{-1}$. In Sect.\,\ref{sec:parametric-plane-omega}, we extend our study over a wide range of the $\Omega_{\textrm{sp}}$--values by exploring the parametric planes. It is also noteworthy that recent works present evidence that the spiral arms closely trace the Milky Way’s rotation curve and exhibit a transient nature (\citealp{2021A&A...652A.162C}; \citealp{2023AJ....166..170J}), and several authors have investigated the growth and disruption phases of the Galactic spiral structure (e.g., \citealp{2025MNRAS.537.2403L}). All of these hypotheses fail to account for the MGs and the DRs because the life-time of the material arms is not long enough to capture the objects in the resonant configurations.

{The amplitude $\zeta_0$ of the spiral arms is determined from global fits of the maximum ratios of radial and tangential forces to the axisymmetric force across the Galactic disk through Eqs.1--4 above. The maximum ratio of forces, as a function of the Galactic radius, lies
in an interval of 3\% to 6\%. These values are consistent with values of density contrast  
collected in the literature by \cite{2011MNRAS.418.1423A} from both Galactic and extragalactic sources. Furthermore, because we employ a four–rigid-body–arm model, the value of $\zeta_0$ is the same for all arms.
It should be noted that, from a dynamical point of view, the value of $\zeta_0$ has no influence on location and modifies the width of the resonance zones only weakly.} 

Finally, $\gamma_{\textrm{sp}}$ is the initial phase angle of the spiral arms, determined by assuming that the Sagittarius arm is located 1\,kpc from the Sun, where the distance of the Sun to the Galactic center is $R_\odot = 8.122$\,kpc and $\varphi_\odot = 90$\,deg. {It should be stressed that this particular orientation of the spiral arms is derived from the observed positions of zero-age masers (see \citealp{2021FrASS...8...48B} and \citealp{2019ApJ...885..131R}); naturally, using different test objects would lead to different inferred orientations of the spiral pattern (e.g., \citealp{2020ApJ...900..186E}, \citealp{2021A&A...651A.104P} and \citealp{2023A&A...670L...7P}).}
 
\begin{figure}
    \includegraphics[width=1.\columnwidth]{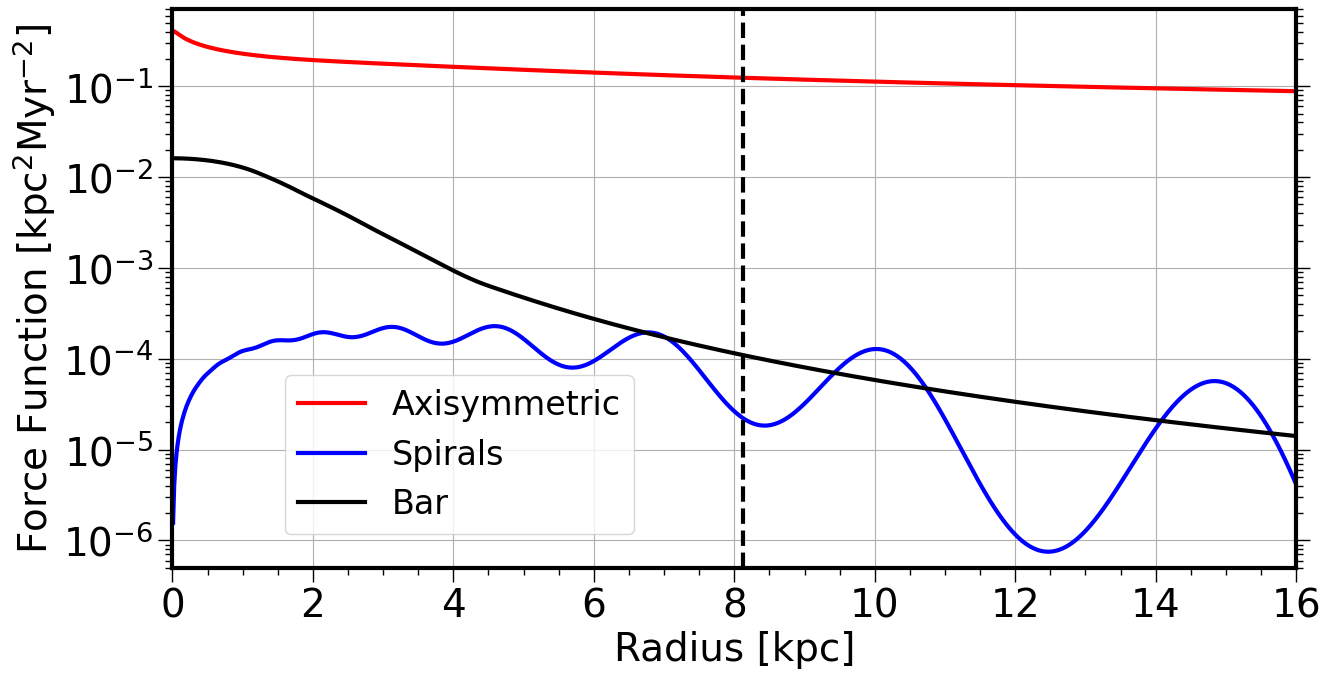}
    \caption{Force functions of the Galactic components: the axisymmetric structure (red), the spiral arms (blue) and the central bar (black), calculated along the radial Galactocentric distance in the direction of the Sun ($\varphi=90$\,deg). The dashed vertical line indicates the actual radial distance of the Sun.}
    \label{fig:force-function}
\end{figure}

\subsection{Central bar's parameters}\label{sec:bar-parameters}

Due to ongoing uncertainties about the physical and structural characteristics of the central bar, we have chosen values that are commonly referenced in recent research in the literature. They are listed in Table\,\ref{tab-init-2}.

\textit{Mass}: Taking a bar-bulge structure with total mass $M_b$, the mass of the bar can be taken as a fraction $\mu = M_{\textrm{bar}}/M_b$ of the total mass, where $\mu$ is a real number between 0 and 1. As an example, \cite{2017MNRAS.465.1621P} obtained the total mass $M_b = 1.88\times10^{10}$\,$M_\odot$ and the resulting bar mass of $M_{\textrm{bar}} = 5.40\times10^{9}$\,$M_\odot$, getting $\mu\approx0.29$. We use the recent value of the total mass fitted in \cite{2023A&A...680A..40M} listed in Table\,\ref{tab-init-2}, and we consider by approximation that the fractional mass of the bar is $\mu=0.30$, which is consistent with the literature. 

\textit{Size}: Several works involving $N$-body simulations support the existence of the long bar structure (e.g., \citealp{2016ARA&A..54..529B} and \citealp{2017MNRAS.465.1621P}), and others show its observational evidence, as pointed by \cite{Zoccali_Valenti_2016}. The estimation of its size varies between 4 and 5 kpc. Therefore, we consider a $R_{\textrm{bar}} = 4.5$\,kpc which is consistent with the values found in the literature (e.g., \citealp{2015MNRAS.450.4050W}, \citealp{2021arXiv211105466T} and \citealp{2023MNRAS.523..991G}), and we also assume a flattening of 0.7, as listed in Table\,\ref{tab-init-2}.    

\textit{Rotation speed}: There are two main scenarios presented in the literature on the value of the bar's rotation pattern speed. The fast rotating bar with $\Omega_{\textrm{bar}} \gtrsim 50.0$\,km\,s$^{-1}$\,kpc$^{-1}$ (e.g., \citealp{2000AJ....119..800D}) is the old one which could explain the origin of HMG by the mechanism of the -2:1 LR. The most recent scenario uses the slow rotating bar with $\Omega_{\textrm{bar}} \lesssim 40.0$\,km\,s$^{-1}$\,kpc$^{-1}$ (e.g., \citealp{2017ApJ...840L...2P}), which is the most acceptable value. The slow rotating bar scenario is supported by arguments that associate the formation of HMG with the CR and high-order LRs (see Sect.\,\ref{sec:topology}). Therefore, we use the value of $\Omega_{\textrm{bar}} = 40.0$\,km\,s$^{-1}$\,kpc$^{-1}$, but we also explore a range of $\Omega_{\textrm{bar}}$--values by using parametric planes in Sect.\,\ref{sec:parametric-plane-omega}.

\subsection{Comparing the force functions of the Galactic components}

A force function, by definition, is a gravitational potential with a positive sign; thus, its larger magnitude will
manifest itself as a stronger perturbation on the stellar motion. In Fig.\,\ref{fig:force-function}, we present the force functions of the Galactic components on a logarithmic scale to facilitate a comparison of their potential contributions. The force of the axisymmetric term (red curve) is dominant in the SNd, at least three orders of magnitude larger than those of the central bar (black curve) and the spiral arms (blue curve). Recall that the axisymmetric potential obtained through Eq.(\ref{eq:PotAxs}), accounts for the contributions of the Galactic disk and the spherical projections of the bulge and dark halo on the Galactic equator. 

The non-axisymmetric components were calculated through Eq.(\ref{eq:PhiArms}), for the spiral arm structure, and Eq.(\ref{eq:PotElipsoide}) (deducting spherical part), for the central bar, along the line of sight from the Sun to the Galactic center ($\varphi = 90$\,deg in the reference frame described below). Close to the Sun's position (vertical dashed line), the forces exerted by the bar and the spiral arms structure are of the same order of magnitude. 

\begin{figure}
    \includegraphics[width=1.\columnwidth]{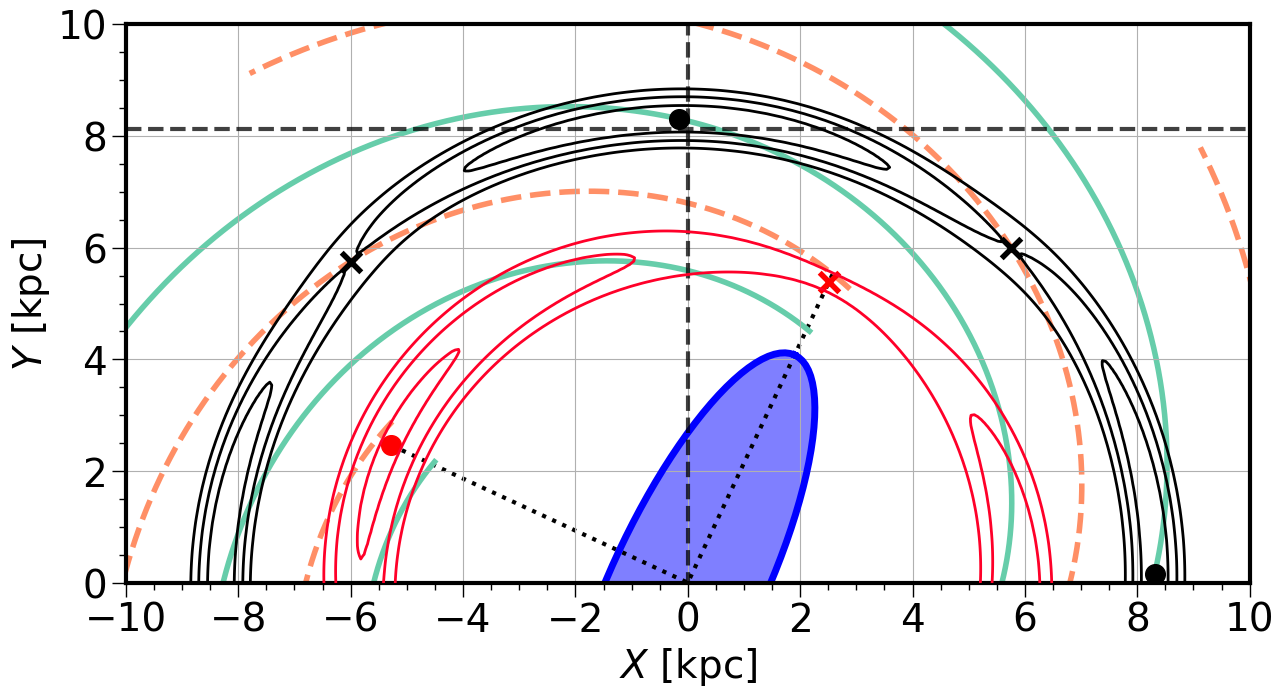}
    \caption{The topology of the Hamiltonian (\ref{eq:Hamiltonian})  on the $X$--$Y$ plane, and the co-rotation zones originated by the spiral arms (black curves) and the central bar perturbations (red curves). The stable and unstable centers (equilibrium solutions) are indicated by the dot and cross symbols, respectively. The black dashed lines intersecting denote the Sun's position, whereas the dotted lines show the orientation of the central bar, which is symbolically depicted as an ellipsoid. We also plot the azimuthal minima and maxima of spiral potential (light green line and orange dashed line, respectively), calculated through Eq.(\ref{eq:PhiArms}). }
    \label{fig:XY_levels}
\end{figure}

\section{Hamiltonian topology, corotation zones and Lindblad resonances}\label{sec:topology}

In the rotating reference frame around the $Z$--axis, the Hamiltonian that describes the 2D stellar dynamics in the equatorial Galactic plane is defined by the axisymmetric potential $\Phi_0(R)$ and the perturbation $\mathcal{H}_1(R,\varphi)$ as
\begin{equation}\label{eq:Hamiltonian}
   \mathcal{H} = \frac{1}{2}\left(p_R^2 + \frac{L_z^2}{R^2}\right) - \Omega_pL_z + \Phi_0(R) + \mathcal{H}_1(R,\varphi),
\end{equation}
where $p_R$ and $L_z$ are canonical momenta, per unit mass, conjugated to $R$ and $\varphi$, respectively, and $\Omega_p$ is a constant angular speed of the frame ({rigid body approximation}). The perturbation $\mathcal{H}_1$ is defined as a sum of the spiral $\Phi_{\textrm{sp}}(R,\varphi)$ and bar's $\Phi_{\textrm{bar}}(R,\varphi)$ potentials: 
\begin{equation}\label{eq:PertHamiltonian}
    \mathcal{H}_1(R,\varphi) = \Phi_\textrm{sp}(R,\varphi) + \Phi_\textrm{bar}(R,\varphi).
\end{equation}

{As shown in Fig.\,\ref{fig:force-function}, the non-axisymmetric terms $\Phi_{\textrm{sp}}(R,\varphi)$ and $\Phi_{\textrm{bar}}(R,\varphi)$ are a few orders of magnitude smaller than the axisymmetric term $\Phi_0(R)$; therefore, they introduce only small perturbations in stellar motions. In perturbation theories, the key assumption is that a complex problem can be treated by successive approximations of a set of simpler related problems. In our case, the first-order approximation consists of the one-of-degree-of-freedom axisymmetric problem, which is perturbed separately by either the central bar or the spiral arms. In the second-order approximation, which accounts for the interaction between two non-axisymmetric structures, the resulting perturbation term becomes considerably weaker, by an order of magnitude, and will be disregarded in this work. Therefore, bar and spiral perturbations will be examined individually in the Hamiltonian (\ref{eq:Hamiltonian}). It should be noted that effects of the interaction of the two patterns  could be of particular interest and will be examined in future work
}.

In Fig.\,\ref{fig:XY_levels}, we analyze the Hamiltonian topology plotting the energy levels of the effective potential (see \citealp{2018ApJ...863L..37M} for details):
\begin{equation}\label{eq:Phi_eff}
    \Phi_{\textrm{eff}}(R,\varphi) = \Phi_0(R) + \Phi_1(R,\varphi) - \frac{1}{2}\Omega_p^2R^2,
\end{equation}
where the perturbation $\Phi_1(R,\varphi)$ is attributed to the central bar or the spiral arms. The set of red curves shows the energy levels, assuming
$\Phi_1(R,\varphi)=\Phi_\textrm{bar}(R,\varphi)$, while the set of black curves shows the energy levels of the choice $\Phi_1(R,\varphi)=\Phi_\textrm{sp}(R,\varphi)$. In the first case, the pattern speed $\Omega_p=\Omega_\textrm{bar}=40.0$\,km\,s$^{-1}$kpc$^{-1}$, while, in the second case,  $\Omega_p=\Omega_\textrm{sp}=28.0$\,km\,s$^{-1}$kpc$^{-1}$ (see Table\,\ref{tab-init-2}). In both cases, the banana-shaped islands, which appear in Fig.\,\ref{fig:XY_levels}, match \textit{corotation zones} on the $X$--$Y$ half-plane (see \citealp{2017A&A...597A..39M} for more details). 

\begin{figure}
    \includegraphics[width=1.\columnwidth]{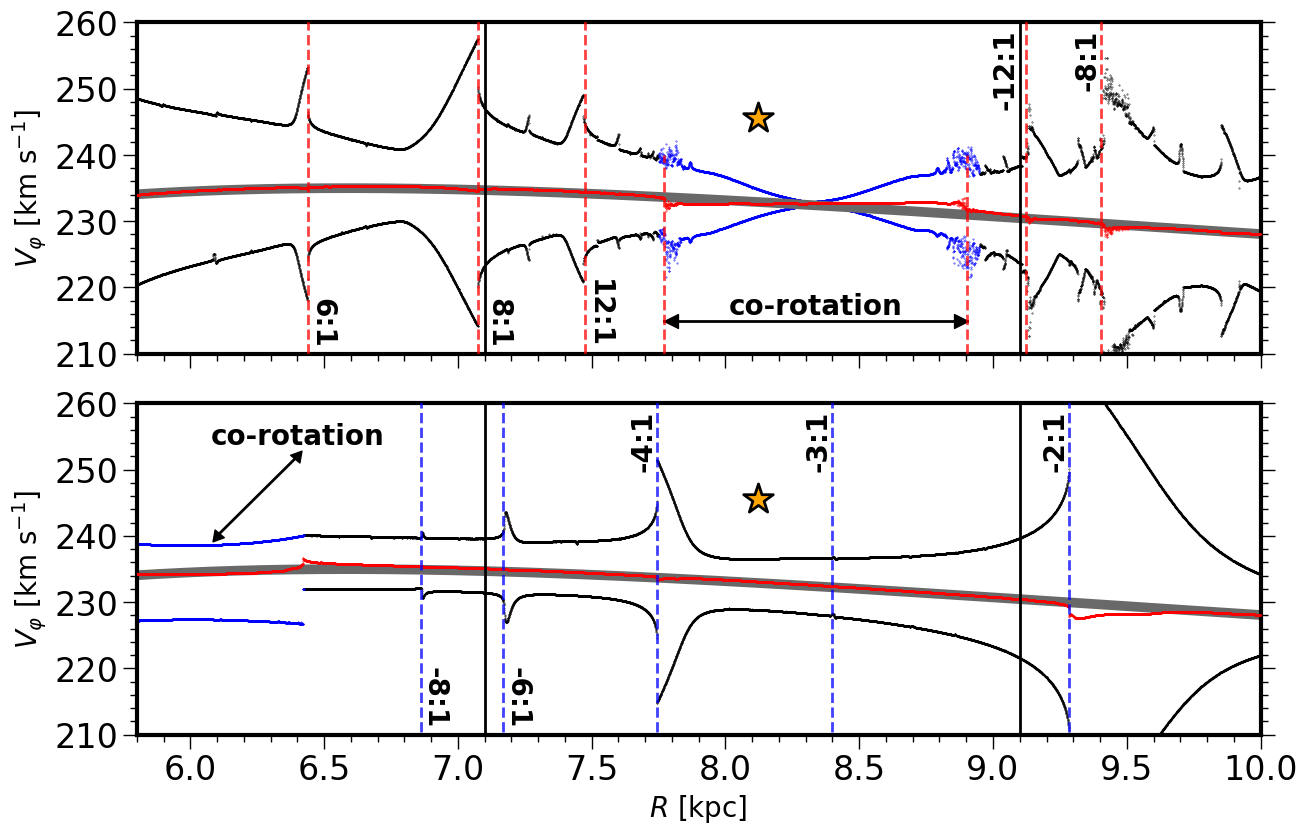}
    \caption{The azimuthal velocity of the initially circular orbits  affected by the spirals (top panel) and the bar (bottom panel) perturbations, as a function of the Galactocentric distance. The maximal/minimal values (black dots) of the velocity $V_\varphi$  were calculated over 10\,Gyr, for $\varphi=90$\,deg. Its averaged values (red dots) is superposed the observation rotation curve (solid gray line). 
    The CR zones on both panels are indicated by the blue dots, while the locations of the main LRs are indicated by vertical dashed lines. The star symbol indicates the initial position of the Sun for reference, while the vertical black lines mark the $R$--interval of completeness as defined in Fig.\,\ref{fig:RLz-distribution}. }
    \label{fig:radial_vel}
\end{figure}

Originating from a \textit{corotation resonance} (CR), corotation zones exhibit diverse attributes, shaped by the particular physical structures that generate them. For instance, the number of the zones depends on the symmetry of the perturber; four islands are formed by the spiral pattern with four arms, while two islands arise from the elongated elliptic structure of the Galactic bar. The location of the zones is defined by the corotation radii of the equilibrium solutions, which are strongly dependent on the value of the pattern speed $\Omega_p$. In our case, the spiral {corotation radius is $8.34$\,kpc }calculated for $\Omega_\textrm{sp}=28.0$\,km\,s$^{-1}$kpc$^{-1}$, while  the bar's {corotation radius is $5.83$\,kpc,} for $\Omega_\textrm{bar}=40.0$\,km\,s$^{-1}$kpc$^{-1}$. Moreover, the orientation of the corotation zones with respect to the Sun on the $X$--$Y$ plane is defined by the phase angles $\gamma_\textrm{bar}$ and $\gamma_\textrm{sp}$, respectively. Finally, the extension of the corotation islands is determined by parameters such as the perturbation amplitude and the total mass, for the spiral structure and the central bar, respectively (see Table\,\ref{tab-init-2}).

\begin{figure*}
    \begin{center}
    \includegraphics[width=2.\columnwidth]{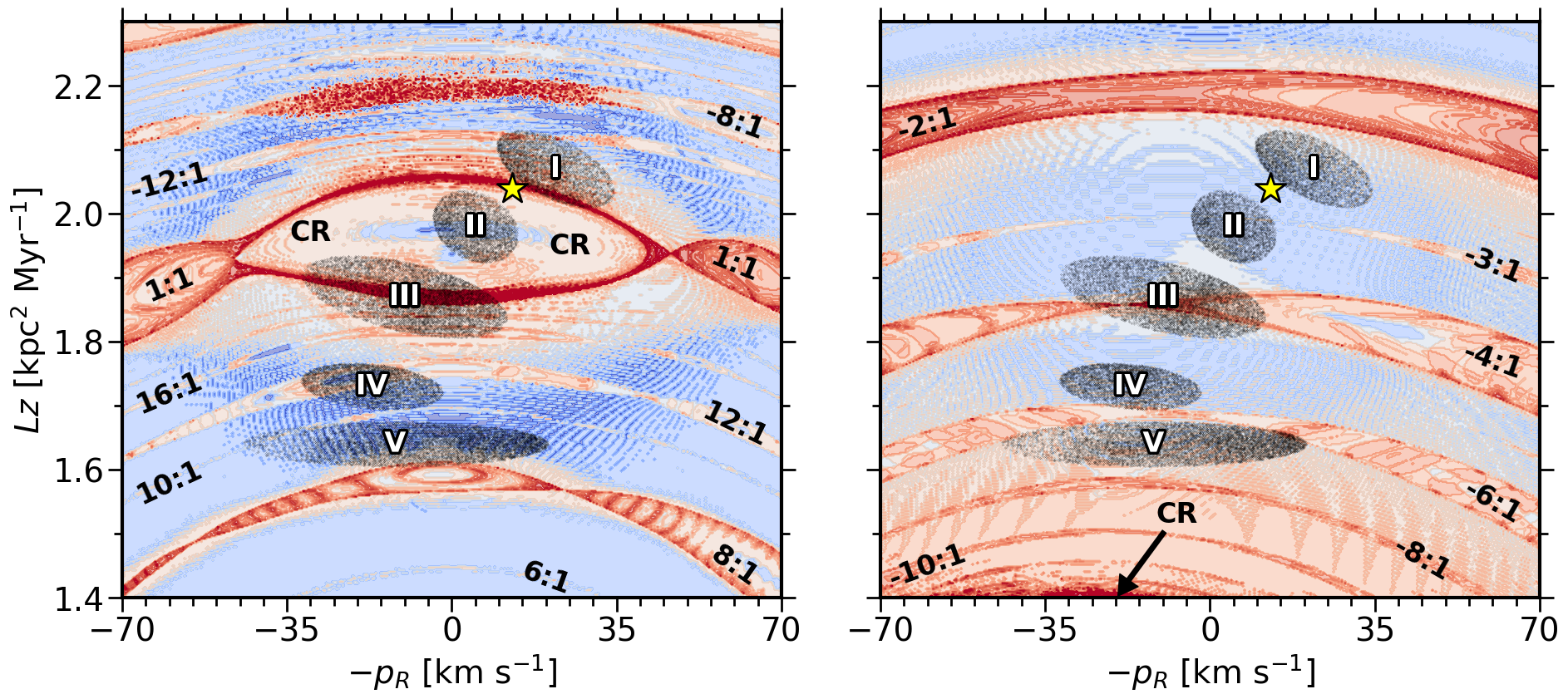}
    \caption{Dynamical maps of the resonances on the $p_R$--$L_z$ plane considering the perturbations from the spiral arms (left panel) and the central bar (right panel). Color scaling enables us to identify the resonances depicted as hot-color (strongly perturbed) structures, while slightly perturbed motions are indicated by cool colors. The main resonances are labeled; the star symbols show the position of the Sun in both planes, for reference.
    The same MGs plotted in Fig.\,\ref{fig:moving-groups} (left panel) are superposed as black dots and identified (I - Sirius, II - Coma Berenice, III - Hyades/Pleiades, IV - HMG I, V - HMG II). } 
    \label{fig:UV-obs-plane}
    \end{center}
\end{figure*}

Figure \ref{fig:radial_vel} shows the effects of gravitational perturbations caused by non-axisymmetric structures of stars originally with circular and undisturbed orbits. The initial positions of these orbits were selected along the analytical rotational curve (solid gray lines on both planes) and propagated over 10\,Gyr, with the initial $p_R=0$ and $\varphi=90$\,deg. The maximal and minimal values of the transversal velocities of stars, $V_\varphi$, are shown by black dots as functions of initial positions, for the spiral arms perturbations (top plane) and bar's perturbations (bottom panel). The average values of $V_\varphi$ are plotted with red dots; they closely align the analytic rotation curve, showing deviations only near the borders of the resonances. 

The main effects of the perturbation observed in Fig.\,\ref{fig:radial_vel} are the sharp increases in the gradual change of $V_\varphi$ relative to $R$. They are originated by Lindblad resonances\footnote{The notation $\pm\,m:n$ represents the LRs, where $m$ and $n$ are simple integers. The positive sign indicates the inner LR, while the negative sign denotes the outer LR.} (LRs), which excite the $V_\varphi$--amplitudes. The main LRs are indicated by vertical dashed lines and are labeled in Fig.\,\ref{fig:radial_vel}. The evolution of the velocity in the corotation zones is indicated by blue dots on both graphs.

\section{Dynamical view of the moving groups...}\label{sec:mapping-resonances}

From a dynamical point of view, CRs and LRs should be responsible for observable features in the disk, such as, for instance, changes in the stellar/gas density distributions. In order to identify the resonances that play a role in forming the MGs, we construct dynamical maps of the SNd and superimpose the star groups on them. 

The construction of the maps is based on the peculiar behavior of the stars orbiting in the resonances, which we observed in Fig.\,\ref{fig:radial_vel}: the smooth evolution of the orbital elements over a grid of initial conditions is sharply changed when a resonance is approached. The spectral analysis technique links this behavior with a chaos indicator, which is visualized on the planes that represent initial conditions (for more details on the spectral analysis method and its applications in the Galactic dynamics studies, we refer the reader to \cite{2002Icar..158..343M} and \cite{2017A&A...597A..39M,2018A&A...615A..10M}, respectively).

In the next, we present the dynamical maps on the $p_R$--$L_z$ and $R$--$L_z$ planes and compare them with the distribution of the observable structures shown in Fig.\,\ref{fig:moving-groups}.

\subsection{... on the $p_R$--$L_z$ plane}\label{sec:pRLz-plane}

Figure \ref{fig:UV-obs-plane} shows two dynamical maps corresponding to the spiral arms model (left) and the central bar model (right). The stellar orbits were calculated on the grids of 401$\times$401 initial conditions, fixing $R = 8.122$\,kpc and $\varphi = 90$\,deg with the parameters from Table\,\ref{tab-init-2}.

The resonances can be observed on both maps, where they are depicted as chains of islands of different widths. The subsets of stars selected from the main MGs in Fig.\,\ref{fig:moving-groups} (left panel) illustrate, by black dots, the positions of the main MGs with respect to the resonances. 

The corotation zones (labeled as CR) appear on both planes in Fig.\,\ref{fig:UV-obs-plane}. The spiral CR (hereafter, SCR; left panel) and its influence zone\footnote{The influence zone of the SCR is composed of overlapping high-order inner and outer SLRs, close to the SCR separatrix.} incorporate three large MGs, Coma-Berenices, Hyades/Pleiades and Sirius. The Sun, whose position is shown by a star symbol, also belongs to the SCR zone. It is worth mentioning that these groups are oriented along the boundary of the spiral corotation zone that can be related with the large amplitudes of the $L_z$--variation within the corotation (see Fig.\,\ref{fig:SOS-spi-resonances} below). Similar slopes are significantly smaller for the two Hercules groups, HMG I and HMG II, aligned along a constant $L_z$-level; their locations are associated with the  12:1 SLR and 8:1 SLR, respectively.

The corotation zone of the bar is located in the smaller $L_z$--values and only its upper border is visible on the dynamical map in Fig.\,\ref{fig:UV-obs-plane} (right panel). However, even though the BCR (CR of the bar) is below the lower boundary of our sample's completeness region, its influence zone reaches as far as the HMG II group. 

All BLRs (bar's LRs) that appear on the map are outer resonances; of these, the most prominent -2:1\,BLR is very close to the upper border of the completeness zone.  We can suggest that the ``Hat"\,feature in the star distribution (see Fig.\,7\,right in \citealp{2025NewAR.10001721H}), may be associated with the overlap of the strong -2:1\,BLR and the -8:1\,SLR. However, the simulations performed by \cite{2020ApJ...888...75B} (which do not contain a bar, only the spiral model), clearly reproduce both the Hercules and Hat features, often used to measure the bar pettern speed.

The two large MGs, Sirius and Coma-Berenices, cannot be associated with any significant BLRs. The Hyades/Pleiades group is crossed by the sufficiently strong -4:1\,BLR. However, the effective area of this resonance around the Hyades/Pleiades group is reduced, because its unstable center corresponds to the observable phase angle of the bar of $115$\,deg. However, it is expected that its overlap with the SCR would produce significant dynamical effects. Finally, two HMG are very close to the -6:1\,BLR.

\begin{figure}
    \includegraphics[width=1.\columnwidth]{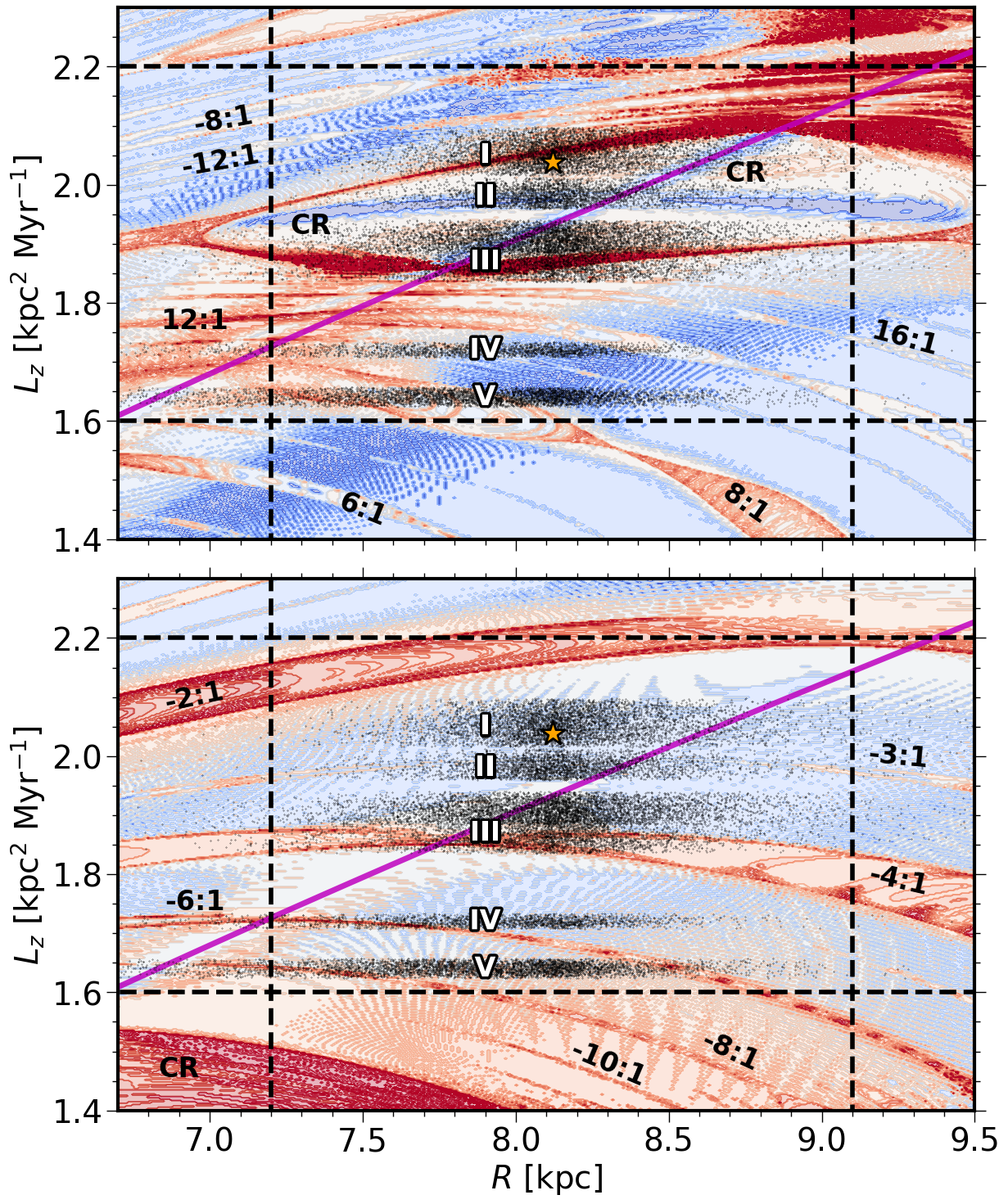}
    \caption{Same as Fig.\,\ref{fig:UV-obs-plane}, except on the $R$--$L_z$ plane. Dynamical map obtained with the spiral arms model is shown on the top panel, while the map corresponding to the central bar model is shown on the bottom panel. The resonances are labeled on both planes; the star symbols show the position of the Sun  on the $R$--$L_z$ plane, for reference, and the continuous purple line represents the observation rotation curve.
    The MGs forming the DRs in Fig.\,\ref{fig:moving-groups} (right panel) are superposed as black dots on the maps. The black dashed lines delimit the region of completeness. }
    \label{fig:RLz-PLANE-OBS-RESONANCES}
\end{figure}

\subsection{... on the $R$--$L_z$ plane}

The maps in Fig.\,\ref{fig:RLz-PLANE-OBS-RESONANCES} show the dynamical structure of the $R$--$L_z$ planes, which result  from the perturbations by the spiral arms (top panel) and the central bar (lower panel). The maps were calculated in a grid with 401$\times$401 initial conditions, using the parameters of Table\,\ref{tab-init-2} and fixing $p_R = 0$ and $\varphi = 90$\,deg. The black dots are subsets of stars of the main MGs, which are superimposed on the dynamical maps.

The resonant zones analyzed previously on the $p_R$--$L_z$ planes can now be identified as chains of islands extending beyond the immediate vicinity of the Sun. The SCR zone produced by the spiral arms (top panel) is a prominent structure near the SNd, along with the inner and outer high-order SLRs, 12:1 and -16:1. Three MGs, such as Sirius, Coma-Berenices and Hyades/Pleiades, are involved in the regions of the SCR and its influence zone.

The HMG II group, a component of the Hercules moving group, appears as a effect of the 8:1\,SLR, while HMG I is more closely aligned with the -6:1\,SLR generated by the central bar (see Fig.\,\ref{fig:RLz-PLANE-OBS-RESONANCES}\,bottom). The Hyades/Pleiades group is clearly affected by the strong -4:1\,BLR of the bar, even if the group's position matches the location of the unstable center of this resonance. The other MGs cannot be associated with the effect of any BLR.

The most prominent bar resonance zones, the BCR and the -2:1\,BLR, are beyond the completeness interval of the observational data, indicating that these resonances cannot explain the dynamical origin of structures close to the SNd. 

\begin{figure}
    \begin{center}
    \includegraphics[width=1.\columnwidth]{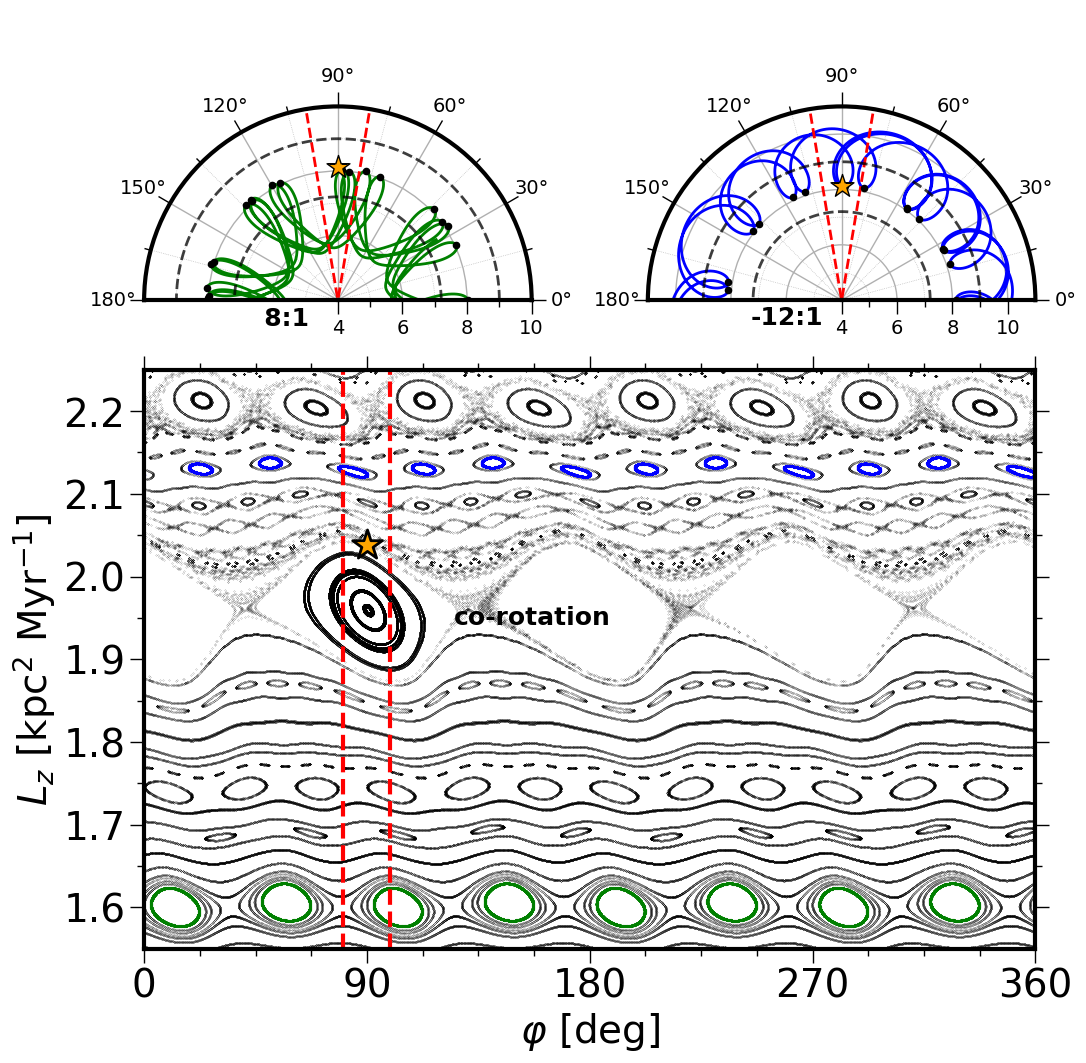}
    \caption{An example of resonant orbits and Poincaré sections obtained with the spiral arms model. Top: The 8:1\,SLR (left panel) and -12:1\,SLR (right panel) orbits in the SNd; the location of the Sun is shown by a star symbol.  The black dots represent the star's transit through the turning point on its orbit.  Bottom: The Poincaré map of the stellar orbits of the same energy on the $\varphi$--$L_z$ plane. The projections are calculated at the instants when the star crosses the plane $p_R = 0$ with $\Dot{p}_R < 0$. The resonant 8:1 and -16:1 orbits appear on the Poincaré map as chains of the green and blue islands, respectively. The dashed red lines delimit the SNd on the $\varphi$--$L_z$ and polar planes. }
    \label{fig:SOS-spi-resonances}
    \end{center}
\end{figure}

\section{Resonant orbits as a possible origin of the moving groups}
\label{sec:origin-groups}

To understand the role of resonances on the formation of MGs, we present the resonant orbits in Figs.\,\ref{fig:SOS-spi-resonances}--\ref{fig:SOS-bar-resonances} corresponding to the spiral arms and bar models, respectively. For better visualization, only segments of the orbits are shown as the stars move across the SNd.

In the case of spiral arms perturbations, we plot in the configuration space two orbits evolving in the SLRs 8:1 (left-top panel) and -12:1 (right-top panel) in Fig.\,\ref{fig:SOS-spi-resonances}. In both cases, we can observe peculiar structures, which are characteristic of near-resonant stellar motions in the rotating framework; they are loops originated by a retrograde motion of stars during their prograde (positive $\dot\varphi$) rotation around the Galactic center. The number of loops per period of an orbit in the {$n$:1} LR is equal to the integer $n$. The loops extend outward from the rotational motion in the 8:1 SLR, while they are oriented inward for the -12:1 SLR. Due to these resonant features, distant stars are able to travel into the SNd.

In a nearly resonant configuration, the stellar orbits exhibit loops slowly precessing around the Galactic center, moving backward for inner LRs and forward for outer LRs. In contrast, within the resonance, the loops oscillate around positions identified as stable stationary points of the resonance. This leads to a concentration of stars in areas associated with resonant stable configurations, thereby increasing the stellar density in these regions and resulting in the formation of the MGs. For example, for one orbit in the 8:1 SLR, the time which the star spends in one loop is approximately 130\,Myr, in one rotation around the Galactic center is 1\,Gyr, while the synodic period of the oscillation of the loop around the resonance center is 5.5\,Gyr. The numerical simulations performed in \cite{2020ApJ...888...75B} showed that the formation of long-lived kinematic structures is the result of stellar orbits trapped by spiral resonances in timespans of the order of 1\,Gyr.

The described dynamical effect of the LRs can be clearly seen on Poincaré map\footnote{The definition and construction of Poincaré maps (Surfaces of Section Method) are described in more detail in \citealp{2018ApJ...863L..37M}} on the bottom panel in Fig.\,\ref{fig:SOS-spi-resonances}. The map was constructed along a fixed energy level, which crosses the SNd, thus providing an overview of the spiral's resonant orbits close to the Sun. 

The map shows on the $\varphi$--$L_z$ plane the orbits traced by the turn points ($p_R$=0) of the loops, which are represented by the black dot symbols in Fig.\,\ref{fig:SOS-spi-resonances}\,top. The chains of islands depicted on the map represent resonant orbits, which could lead to the formation of the observed MGs within the SNd. The SLRs are easily identified by the number of islands, which represent the resonant loops. In contrast, the non-resonant orbits appear as sinuous patterns, becoming more pronounced near one of the SLRs. Resonant stars spend their orbital time gathering inside the islands, forming observable groups. The SCR dominates the map for the chosen $\Omega_{\textrm{sp}}=28.0$\,km\,s$^{-1}$\,kpc$^{-1}$, the 8:1\,SLR and the -12:1\,SLR appear as green and blue chains of islands, respectively. 

Figure \ref{fig:SOS-bar-resonances} shows two resonant orbits and the Poincaré map calculated with the central bar model.
The orbits are evolving in the resonances -6:1 (left top panel) and -4:1 (right top panel), and the motion of their turn points on the Poincaré map (bottom panel) is plotted by the green and blue chains of islands, respectively.  When comparing the resonant orbits and dynamical structures of the maps for both the spiral arm and bar models, we can observe notable qualitative similarities, with differences mainly in quantitative characteristics. Thus, the main effects of small perturbations that originated from a non-axisymmetric structure on the axisymmetric disk are resonances whose symmetry, positions, and amplitudes will depend on the chosen parameters.

The vertical component of the angular momentum, $L_z$, is fundamental in examining the effect of perturbations on the orbits of stars. In fact, in an unperturbed problem, $L_z$ is an integral of stellar motion. In the perturbed, but non-resonant case, the $L_z$--variations are small that can be observed on the Poincaré maps as slight undulations of the nearly constant $L_z$--values (Figs.\,\ref{fig:SOS-spi-resonances}--\ref{fig:SOS-bar-resonances}). When approaching a resonance, the amplitude of the $L_z$ oscillation increases, finally achieving its peak values at the resonant state. The variation of $L_z$ may be used as a resonance measurement: elongations of HMG I and II in relation to the constant $L_z$ could be linked to high-order LRs, while the significant slopes of the groups Sirius, Coma-Berenices and Hyades/Pleiades in the momentum space are influenced by SCR induced by the spiral arm.

\begin{figure}
    \begin{center}
    \includegraphics[width=1.\columnwidth]{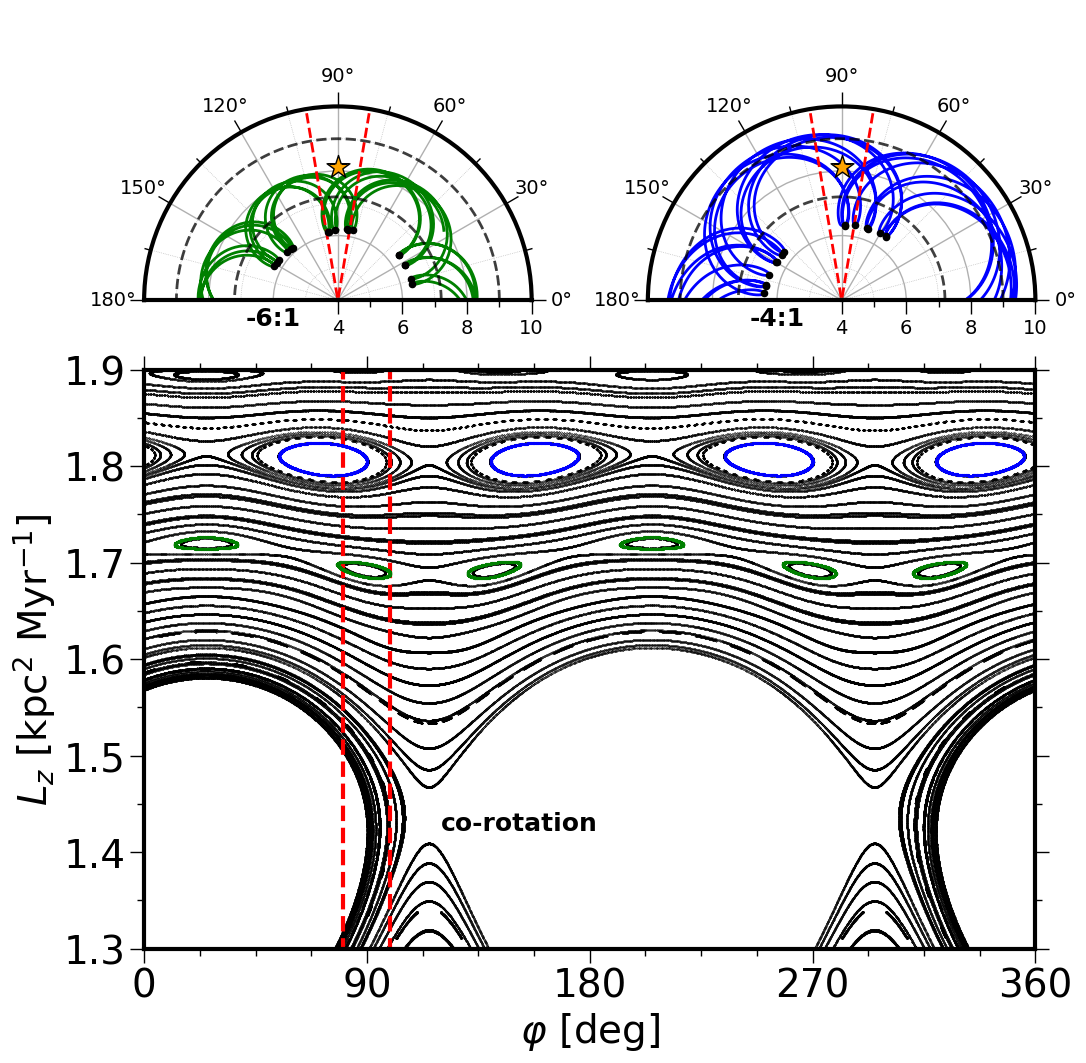}
    \caption{Same as Fig.\,\ref{fig:SOS-spi-resonances}, except obtained with the central bar model. The orbits on the top panels are -6:1\,BLR (left) and -4:1\,BLR (right) orbits, which appear as chains of green and blue islands, respectively, on the Poincaré map (bottom panel). }
    \label{fig:SOS-bar-resonances}
    \end{center}
\end{figure}

\begin{figure}
    \begin{center}
    \includegraphics[width=1.\columnwidth]{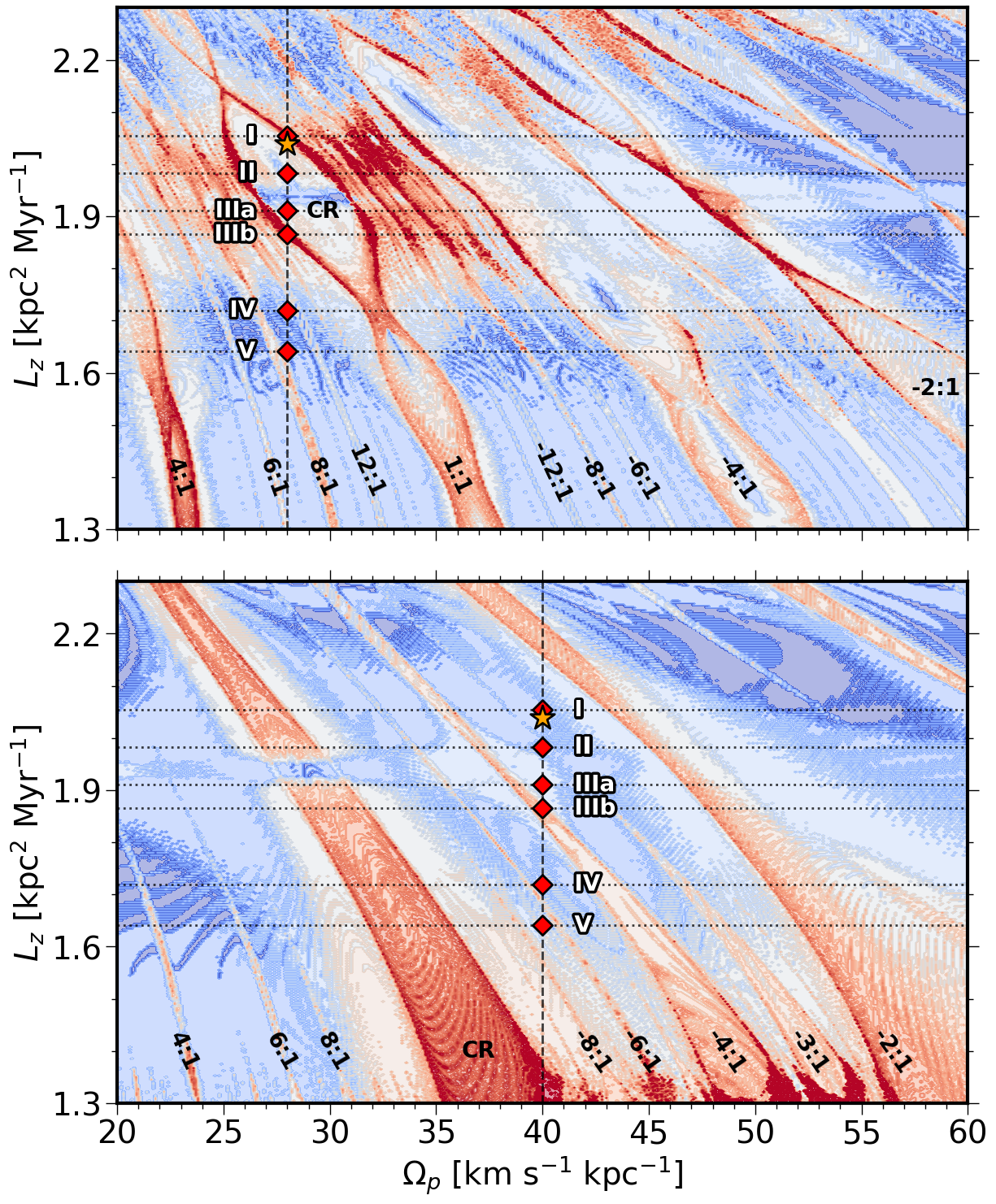}
    \caption{Dynamical maps on the parametric $\Omega_p$--$L_z$ planes calculated for the spiral arms (top) e central bar (bottom) perturbations on $401\times401$ grids, fixing $R=R_\odot$, $p_R = 0$ and $\varphi = 90$\,deg.
    The resonance locations are identified by the corresponding labels; the star symbols indicate the position of the Sun on both planes. The diamond symbols show schematically positions of the main MGs, from I to V,  obtained for  $\Omega_p$ equal to 28\,km\,s$^{-1}$\,kpc$^{-1}$ (top) and 40\,km\,s$^{-1}$\,kpc$^{-1}$ (bottom).}
    \label{fig:parametric-plane-omega}
    \end{center}
\end{figure}

\section{Parametric planes}
\label{sec:parametric-plane-omega}

In order to evaluate a resonance membership of the sample objects, we applied our model for the spiral arms and the central bar, separately. The behavior of stars from the core of each group was analyzed, and the result shows good agreement with the theoretical predictions presented above. The HMG I, with 5400 members, contains 25.0\% of the population in the 12:1\,SLR and 5.0\% in the 10:1\,SLR, while only 4.8\% is involved inside the -6:1\,BLR. The HMG II, with 6262 members, contains only 6.4\% in the 8:1\,SLR and 1.4\% in the -6:1\,BLR. The members of the Pleiades group are divided between the SCR (46.0\%) and the -4:1\,BLR (48.0\%). The majority of the Hyades population evolves in the SCR zone, while 18.0\% of the members are within the -4:1\,BLR. The Coma-Berenices is deeply evolved within the SCR ({87.0\%}). Finally, the Sirius group is filled by the SCR members (25.0\%) and the rest is evolving inside the high-order SLRs, which are dense near the SCR. However, it should be stressed that these results depend on the applied model and, especially, on the chosen parameters of the model. 

From a dynamical point of view, the choice of pattern rotation speed $\Omega_p$ is relevant, since it defines the locations of resonance zones in the phase space. 
However, both qualitative and quantitative estimations of the $\Omega_p$ values for spiral and bar structures still vary in large ranges (see Fig.\,10 in \citealp{2025arXiv250703742K}).  
In contrast, the angular momentum of stars $L_z$ is a high-precision observable quantity. In this context, we can qualitatively evaluate the parameter $\Omega_p$ utilizing parametric planes that enable us to identify the correlation between the MGs, viewed in terms of $L_z$, and $\Omega_p$.

Figure \ref{fig:parametric-plane-omega} shows two parametric planes $\Omega_p$--$L_z$, constructed for the spiral arms (top panel) and the central bar models (lower panel). The maps were calculated on the grids of 401$\times$401 initial conditions, with fixed $R=R_\odot$, $p_R=0$ and $\varphi=90$\,deg. The red diamond symbols schematically indicate the positions of the main MGs\footnote{To assist in the analysis, the Hyades/Pleiades group is separated in IIIa (Hyades) and IIIb (Pleiades) sub-groups.}, fixed for the values $\Omega_\textrm{sp}=28.0$\,km\,s$^{-1}$\,kpc$^{-1}$ and $\Omega_\textrm{bar}=40.0$\,km\,s$^{-1}$\,kpc$^{-1}$ (see Table\,\ref{tab-init-2}). The main LRs, which appear on the parametric maps, are identified by the corresponding ratios. 

In their current positions, the Sirius (I) and Coma-Berenices (II) groups are found within the zone of influence of the SCR (top panel in Fig.\,\ref{fig:parametric-plane-omega}), but, when analyzed on the parametric plane of the bar (bottom panel), it is observed that there are no resonances that could exert influence on these groups. However, the Hyades/Pleiades group presents an interesting result. By subdividing into two groups, Hyades (IIIa) and Pleiades (IIIb), it is noted that group IIIa is within the SCR, while group IIIb fits within the -4:1\,BLR. This suggests that the Hyades/Pleiades group is probably formed by the overlap of SCR with -4:1\,BLR, which could explain this bimodality observed in Fig.\,\ref{fig:moving-groups}.

Looking specifically at the Hercules group in both planes in Fig.\,\ref{fig:parametric-plane-omega}, it is observed that HMG I (group IV) is very close to 12:1\,SLR and -6:1\,BLR, while HMG II (group V) is close to 8:1\,SLR and -8:1\,BLR. However, according to our numerical tests, only a small fraction of these stars are actually captured in the resonances. This is in part due to the density of the Hercules group being higher at smaller radial distances, as shown in \cite{2018ApJ...863L..37M}. 

Examining Fig.\,\ref{fig:parametric-plane-omega}, we observe that the positions of the LRs on both planes are similar, since they are mainly determined by the chosen value of $\Omega_p$. In this way, the best-fit estimate for $\Omega_p$ presented in \cite{2025arXiv250703742K} could correspond to the central bar or the spiral arms, each rotating at an angular velocity of 28.2$\pm$0.1\,km\,s$^{-1}$\,kpc$^{-1}$.
However, we can also note in Fig.\,\ref{fig:parametric-plane-omega} that the forms and intensities of the LRs are significantly influenced by the model used to represent the non-axisymmetric components of the Galactic potential. Thus, comparing two planes, we conclude that the resonance zones of the spirals align more closely with the MGs positions, despite the spiral force function being less intense than that of the central bar in the region of the SNd (see Fig.\,\ref{fig:force-function}).


\section{Conclusions}
\label{conclus}

This study analyzes the dynamical effects on the stellar motion resulting from the Galactic non-axisymmetric components.  
The resonant features in the SNd were derived using the 2D models for the spiral arms and the central bar perturbations separately. The physical and structural parameters of the models were chosen on the basis of what is typically used in the existing literature. For comparison purposes, the MGs and DRs were selected as test structures with observable properties. We established a completeness limit for the data sample at about 1\,kpc from the Sun, allowing us to compare theoretical predictions with observations while minimizing bias.

Investigating the behavior of stars within the Galactic potential, we constructed maps of the velocity and action spaces in the SNd. We identified the important dynamical characteristics on the representative planes,  the corotation and Lindblad resonances. For the chosen rotation velocity of the central bar, only outer LRs arise in the SNd, whereas for the spiral arms, both inner and outer LRs can be found. To establish the correlation between the resonances and the main MGs, these were superimposed on the dynamical maps. Our findings suggest that the Sirius, Coma-Berenices, and Hyades/Pleiades groups are connected to the SCR, while some portion of the Pleiades group undergoes evolution in the -4:1\,BLR.

The two components of the Hercules group  were associated with the high-order LRs, the inner 8:1 and 12:1 SLRs, and the outer -8:1 and -6:1 BLRs. Given that various studies exploring the HMG's chemical composition (e.g., \citealp{2010LNEA....4...13B}, \citealp{2023ApJ...956..146L}, and \citealp{2025MNRAS.538.1963L}) suggest that the group members migrated from the central part of the Galactic disk, it can be inferred that the inner LRs of spiral arms may be responsible for the transport of stars of the HMG.

A detailed examination of the resonant behavior of the group members allowed us to elucidate the mechanism of the formation of the moving groups. These stars are trapped in loops of resonant orbits, which oscillate around stable periodic configurations, with periods several orders of magnitude longer than the orbital periods around the Galactic center. 

We have demonstrated that the locations of the LRs are strongly influenced by the pattern rotation speed. Given the wide range of estimates from the literature for the pattern speeds of the central bar and spiral arms, we explored parametric planes by adjusting these values with the positions of the main moving groups. Our study suggests that the spiral arms model, with the parameter $\Omega_{\textrm{sp}}=$ 28.0\,km\,s$^{-1}$\,kpc$^{-1}$, more accurately fits the locations of the MGs than the bar model with any pattern speed between 20 and 60\,km\,s$^{-1}$\,kpc$^{-1}$. In addition to positions in velocity and phase spaces, the shapes and stability of the resonant regions formed by spiral arms provide the closest match with the observed MGs.

\section*{Acknowledgements}

We thank our anonymous reviewer for his helpful comments and suggestions. We also thank Dr. Phillip Galli for the useful suggestions concerning the analysis of the Gaia's astrometry data and Dr. Wilton Dias for the valuable insights on the nature of the Galactic spiral arms. This work was supported by the Brazilian funding agencies CNPq, FAPESP and CAPES. This work has made use of data from the European Space Agency (ESA) mission Gaia (\url{https://www.cosmos.esa.int/gaia}), processed by Gaia Data Processing and Analysis Consortium (DPAC, \url{https://www.cosmos.esa.int/web/gaia/dpac/consortium}). This work has made use of the facilities of the Laboratory of Astroinformatics (IAG/USP, NAT/Unicsul), funded by FAPESP (grant 2009/54006-4) and INCT-A.





\bibliographystyle{apj}
\bibliography{references}

\end{document}